\documentclass[10pt,english,aps,prl,showpacs,twocolumn]{revtex4} 
 \usepackage{amsmath} 
 \usepackage{amssymb} 
 \usepackage{graphicx} 
 \usepackage{epsfig} 
 \usepackage{psfig} 
 
\begin{document} 
 
 \title{Superconductivity in charge Kondo systems} 
 \author{ Maxim Dzero and J\"{o}rg Schmalian$\ $} 
 \affiliation{Department of Physics and Astronomy and Ames Laboratory, Iowa State 
 University, Ames, IA 50011, USA } 
 \date{\today} 
 
 \begin{abstract} 
 We present a theory for superconductivity and charge Kondo fluctuations, 
 i.e. resonant quantum valence fluctuations by two charge units, for \textrm{%
 Tl}-doped \textrm{PbTe.} We show that \textrm{Tl }is very special as it 
 first supplies a certain amount of charge carriers \ to the \textrm{PbTe}%
 -valence band and then puts itself into a self-tuned resonant state to yield 
 a new, robust pairing mechanism for these carriers. 
 \end{abstract} 
 
 \pacs{74.10.+v, 74.70.-b, 75.20.Hr} 
 \maketitle 
 
 The role of impurities in superconductors is a classic problem in condensed 
 matter physics\cite{Anderson,Abrikosov60}. A reciprocal problem concerns 
 impurities which can cause superconductivity in a host that, on its own, has 
 no intention to superconduct. One version is of course an impurity induced 
 increase in the carrier concentration and density of states at the Fermi 
 level. Much more exotic and interesting is however the prospect of 
 impurities supplying the actual pairing mechanism. Candidates are so called 
 negative-$U$ centers\cite{Anderson75}, which can, as we will show, induce 
 pairing in a non-superconducting host even in a regime of strong quantum, 
 charge Kondo, fluctuations. \ The latter is crucial to understand  
 superconductivity in \textrm{Pb}$_{1-x}\mathrm{Tl}_{x}$%
 \textrm{Te} \cite{PbTesc}, where recent experiments by Matsushita \emph{et 
 al.}\cite{Fisher04} found strong evidence for charge Kondo fluctuations 
 close to$~T_{c}$. It promises a number of new unconventional properties\cite%
 {Oganesyan02} for this very exciting material. 
 
 \textrm{Pb}$\mathrm{Te}$ is a narrow gap IV-VI semiconductor\cite{Nimtz83} 
 where $\mathrm{Tl}$, for small $x$, is known to act as acceptor, adding one 
 hole per atom to the valence band. This is consistent with the valence 
 electron configurations of \textrm{Pb} ($6s^{2}6p^{2}$) and \textrm{Tl} ($%
 6s^{2}6p^{1}$). The surprise is that \textrm{Pb}$_{1-x}\mathrm{Tl}_{x}$%
 \textrm{Te }becomes superconducting with $T_{c}$ as big as $1.4\mathrm{K}$%
 \cite{PbTesc}, comparable to metallic systems, but for a hole concentration 
 orders of magnitude smaller ($n_{0}\simeq 10^{20}$cm$^{-3}$). Equally 
 puzzling is that $T_{c}$ rises with \textrm{Tl }concentration, $x$, for $x$%
 -values where $n_{0}$ becomes independent of $x$\cite{Murakami96,Nemov98}. 
 
 A special aspect of $\mathrm{Tl}$ is that it likes to skip an intermediate 
 valence state in a polarizable host\cite{Weiser81,Varma88}.
 In \textrm{Pb}$\mathrm{Te}$, $\mathrm{Tl}^{+}$, 
 which acts as an acceptor, and $\mathrm{Tl}^{3+}$, where an electron 
 is donated instead, are by several $\mathrm{eV}$ more stable than 
 $\mathrm{Tl}^{2+}$\cite{Weiser81}. 
 This effect can be described in terms a negative-$U$ Hubbard 
 interaction between holes in the \textrm{Tl}$6s$-shell. If $\delta E=E\left( 
 \mathrm{Tl}^{3+}\right) -E\left( \mathrm{Tl}^{+}\right) $ is the smallest 
 scale of the problem, the two valence states become essentially degenerate. 
 Then, the hybridization of the impurities with valence holes causes a 
 quantum charge dynamics, similar in nature to the Kondo effect of diluted 
 paramagnetic impurities in metals\cite{Hewson,Tarapher91}. An isospin can be 
 introduced\cite{Tarapher91} where the "up" and "down" configurations 
 correspond to $\mathrm{Tl}^{3+}$ and $\mathrm{Tl}^{+}$, respectively.$\ 
 \delta E\neq 0$ plays the role of the magnetic field and the isospin flip 
 corresponds to a coherent motion of an electron pair into or out of the 
 impurity. This motion of pairs suggest a connection between the charge Kondo 
 dynamics, with Kondo temperature $T_{K}$, and superconductivity. Numerical 
 simulations\cite{Schuettler89} indeed demonstrate that negative-$U$ centers 
 increase $T_{c}$ of a superconducting host if $\delta E$ is small. For $%
 \delta E=0$ pairing in a non-superconducting host was discussed under the 
 assumptions $T_{c}\gg T_{K}$\cite{Malshukov91}. 
\begin{figure}
\hbox{(a)\hspace{65mm}(b)}  
\vspace{3mm}
\centerline{
  \hbox{
    \resizebox{49mm}{50mm}
    {\includegraphics{fig1a.eps}}
    \hspace{2mm}
    \resizebox{32mm}{!}
    {\includegraphics{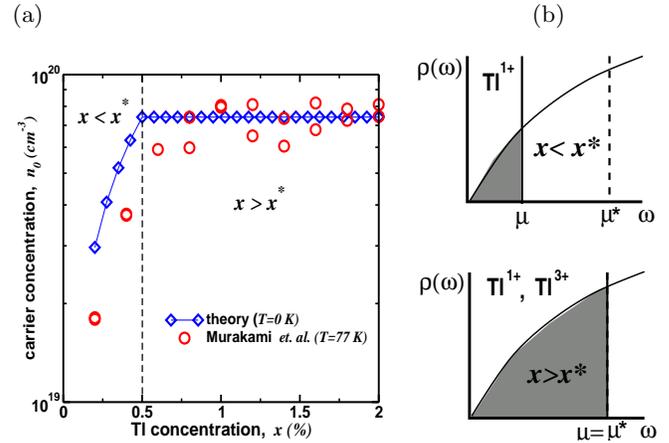}}
  }
}
\caption{(a) Valence band hole concentration as function of Tl content in Pb$_{1-x}$Tl$_x$Te
in comparison with experiment[8]; 
(b) valence band density of states for $x<x^{*}$ (upper panel) and
$x>x^{*}$ (lower panel). A pinning of the chemical potential at $\mu=\mu^{*}$ 
for $x>x^{*}$ gives rise to a degeneracy between 
the Tl$^{1+}$ and Tl$^{3+}$ states and to $n_0(x)=const$.}
\label{Fig1}
\end{figure}

 Two important open questions arise: i) Why is it possible to assume almost 
 perfect degeneracy ($\delta E<T_{c}$) given that $\mathrm{Tl}$ is known to 
 act as acceptor (requiring $E\left( \mathrm{Tl}^{1+}\right) <E\left( \mathrm{%
 Tl}^{3+}\right)$) even at room temperature? ii) Are charge Kondo 
 impurities able to cause superconductivity with $T_{c}\simeq {T_{K}}$, as 
 requires by recent experiments\cite{Fisher04}? Then the scattering rate of 
 the centers is highly singular and the pseudo-spin moment is about to be quenched. 
 
 In this paper we answer both questions. We show that beyond a 
 characteristic \textrm{Tl}-concentration \textrm{Pb}$_{1-x}\mathrm{Tl}_{x}$%
 \textrm{Te} tunes itself, without adjustment of parameters, into a resonant 
 state with $\delta E=0$. We further present a theory for the 
 superconducting transition temperature of dilute negative-$U$, charge Kondo 
 impurities to address the behavior in the intermediate regime $T_{c}\simeq 
 T_{K}$, where the superconducting and charge Kondo dynamics fluctuate on the 
 same time scale. We argue that our theory can explain the concentration 
 dependence and magnitude of $\ n_{0}$ and $T_{c}$ for \textrm{Pb}$_{1-x}%
 \mathrm{Tl}_{x}$\textrm{Te}. In addition we predict a re-entrance normal 
 state behavior at low temperature and impurity concentration as a unique 
 fingerprint of the charge Kondo mechanism for superconductivity, determine 
 the electromagnetic response close to the transition and show that a low 
 concentration of negative-$U$ centers will \emph{always} increase weak 
 coupling host superconductivity. All this demonstrates the rich and highly 
 nontrivial behavior of this very special class of impurities. 
 
 An isolated valence skipper can be described in terms of a negative-$U$ 
 Hubbard model, 
 \begin{equation} 
 H_{\mathrm{imp}}=\left( \varepsilon _{0}-\mu \right) \sum_{\sigma 
 }n_{s,\sigma }+Un_{s\uparrow }n_{s\downarrow } 
 \end{equation}%
 where $n_{s,\sigma }=s_{\sigma }^{\dagger }s_{\sigma }$ is the occupation 
 for a spin $\sigma $ hole in the \textrm{Tl} $6s$-shell, i.e. $\delta 
 E=2\left( \varepsilon _{0}-\mu \right) +U$. $\mu $ is the chemical potential 
 of the system and $U<0$. The valence band is characterized by $H_{\mathrm{%
 band}}=\sum_{\mathbf{k,}\sigma }\left( \varepsilon _{\mathbf{k}}-\mu \right) 
 c_{\mathbf{k\sigma }}^{\dagger }c_{\mathbf{k}\sigma }.$The concentration of 
 holes in the valence band, donated via \textrm{Tl}-doping, is $n_{0}\ 
 =x\left( 1-n_{s}\right) $ with $n_{s}=\sum_{\sigma }\left\langle n_{s,\sigma 
 }\right\rangle $, i.e. $n_{0}>0$ in case of \ an acceptor, $\mathrm{Tl}^{+}$%
 , and $n_{0}<0$ (corresponding to electrons in the conduction band) for the 
 donor, $\mathrm{Tl}^{3+}$. This enables us to determine $\mu $ and thus $%
 \delta E$ as function of \textrm{Tl} concentration. We first assume that the 
 chemical potential, $\mu $, is below the value $\mu^{*}=\varepsilon _{0}+%
 \frac{1}{2}U$, where $\delta E=0$. Then $\delta E>0$ and \textrm{Tl}$^{+}$ 
 is more stable. There are no holes in the \textrm{Tl} $6s$ levels. All holes 
 are in the valence band: $n_{0}=x$, as seen in experiment for small $x$\cite%
 {Murakami96,Nemov98}. Increasing the \textrm{Tl} concentration increases $%
 \mu $ until it reaches $\mu^{*}$ for some $x^{\ast }$. If we further add 
 \textrm{Tl}-impurities and if they continued acting as acceptors, the 
 chemical potential would rise above $\mu^{*}$. However, then $\delta E<0$ 
 and \textrm{Tl}$^{3+}$ become more stable acting as donor, in contradiction 
 to our assumption. Thus, instead of increasing $\mu $, additional impurities 
 will equally split into \textrm{Tl}$^{+}$ and \textrm{Tl}$^{3+}$ valence 
 states such that no new charge carriers are added to the valence band and $%
 \mu $ remains equal to $\mu^{*}$. \textrm{Tl}$^{+}$ and \textrm{Tl}$^{3+}$ 
 are degenerate and coexist with concentration $\frac{x+x^{\ast }}{2}$ and $%
 \frac{x-x^{\ast }}{2}$, respectively. No fine tuning is needed to reach a 
 state with perfect degeneracy, except for the fact that $\mu^{*}$ is 
 reachable. This phenomenon is related, but not identical, to the pinning of 
 the Fermi level in amorphous semiconductors, discussed in Ref.\cite%
 {Anderson75}. \ In Fig.1 we show experimental results of Ref.\cite%
 {Murakami96} for $n_{0}\left( x\right) $, in good agreement with this 
 scenario. The comparison with experiment gives an estimate of 
$x^{\ast }\simeq 0.5\%$ (see Fig. 1). Using the band structure of 
 \textrm{PbTe}\cite{bandstructure} this yields 
 $\mu^{*}\simeq{175\pm{20}}\mathrm{meV}$ 
 and $\mu^{*}\rho _{0}\simeq 0.07$ with density of states at the Fermi 
 level, $\rho _{0}$. This value for $\mu^{*}$ agrees very well with the 
 tunneling data of Ref.\cite{Murakami96}, who finds $\mu^{*}\approx{200}$meV. 
 
 Next we include an additional hybridization of the impurity with the band 
 electrons, $V\sum_{i\sigma }\left( s_{i\sigma }^{\dagger }c_{i\sigma 
 }+c_{i\sigma }^{\dagger }s_{i\sigma }\right) $, \ causing transitions 
 between the degenerate valence states. For large $|U|/V$, the problem can be 
 simplified by projecting out states with $n_{is\sigma }=1$\cite{Schrieffer66}%
 . The close relation to the spin Kondo problem becomes evident if one 
 introduces the Nambu spinor\cite{Tarapher91} $\widehat{c}_{i}=\left( 
 c_{i\downarrow },c_{i\uparrow }^{\dagger }\right) $ as well as the isospin $%
 \mathbf{t}_{i}=\frac{1}{2}\widehat{c}_{i}^{\dagger }\mbox{\boldmath$\tau$}%
 \widehat{c}_{i}$ and similarly $\widehat{s}_{i}$ and $\mathbf{T}_{i}=\frac{1%
 }{2}\widehat{s}_{i}^{\dagger }\mbox{\boldmath$\tau$}\widehat{s}_{i}$. Here $%
 \mbox{\boldmath$\tau$}$ is the vector of the Pauli matrices. For $\delta E=0$ 
 follows 
 \begin{equation} 
 H_{\mathrm{int}}=J\sum_{i}\mathbf{T}_{i}\cdot \mathbf{t}_{i}, 
 \end{equation}%
 where $J=\frac{8V^{2}}{\left\vert U\right\vert }$. The isospins $\mathbf{T}%
 _{i}$ and $\mathbf{t}_{i}$ obey the usual spin commutation relation. 
 Ordering in the $x$-$y$ plane in isospin space is related to 
 superconductivity ($T_{i}^{+}=s_{i\downarrow }^{\dagger }s_{i\uparrow 
 }^{\dagger }$) , whereas ordering in the $z$-direction corresponds to charge 
 ordering ($T_{i}^{z}=\frac{1}{2}\left( \sum_{\sigma }n_{is\sigma }-1\right) $%
 ). The model undergoes a Kondo effect where the low temperature bound state 
 is a resonance \ of a pair of charges tunneling between the impurity and the 
 conduction electron states at a rate $T_{K}\simeq De^{-\frac{1}{\rho _{F}J}}$%
 , forming unitary scattering centers at $T\ll {T}_{K}$ ($D$ is the valence 
 hole band width of order $\mu^{*}$). The analogy to the spin Kondo problem 
 is however not perfect. The valence band part of the Hamiltonian, $H_{\mathrm{band%
 }}=\sum_{\mathbf{k,\sigma }}\left( \varepsilon _{\mathbf{k}}-\mu \right) 
 \widehat{c}_{\mathbf{k}}^{\dagger }\tau _{z}\widehat{c}_{\mathbf{k}}$, is 
 not isospin rotation invariant. This causes an anisotropy of the analog of 
 the RKKY interaction between isospins mediated by either particle-particle 
 excitations, $I^{+-}\left( R\right) =\frac{J^{2}\rho _{F}}{8\pi }R^{-3}$ \ 
 or particle-hole excitations, $I^{zz}\left( R\right) =$ $I^{+-}\left( 
 R\right) \cos \left( 2k_{F}R\right) $, respectively. The in-plane coupling 
 in isospin space, $I^{+-}$, is the Josephson or proximity coupling between 
 distinct impurities, whereas $I^{zz}$ determines charge ordering. The 
 absence of Friedel oscillations in the particle-particle channel  causes the 
 different behavior of $I^{+-}$ and $I^{zz}$. 
 
 Using this pseudospin analogy one can easily conclude that superconductivity 
 is possible if $T_{c}$ turns out to be large compared to $T_{K}$ and quantum 
 fluctuations of $\mathbf{T}_{i}$\ can be neglected. 
 The pseudospin moment is unscreened, corresponding to preformed pairs.
 The interaction $I^{+-}$ between these pairs
 in the $\ $isospin $x$-$y$ plane is unfrustrated, supporting superconducting 
 rather than charge ordering for randomly placed impurities. A mean field 
 calculation in this regime gives $T_{c,\mathrm{mf}}\simeq xJ^{2}\rho 
 _{F}\log \left( D/\left( xJ^{2}\rho _{F}\right) \right)$
 \cite{Malshukov91}. The origin of superconductivity is then similar 
 to Josephson coupling between small superconducting grains located at the 
 impurity sites. 

 For $T_{c}$ comparable to $T_{K}$ the behavior is considerably more subtle. 
 The time it takes to create a Cooper pair in the host equals the time for a 
 valence change causing the pairing, i.e. the moments which are supposed 
 to order are being quenched and a description in terms of preformed pairs is
 inapplicable. In addition, Kondo flip-scattering is expected 
 to be pair breaking. 
 
 Theoretically, the Kondo effect manifests itself in the appearance of the 
 logarithmic divergence of the perturbation theory in $J$ for $T\simeq T_{K}$%
 . A partial summation of the divergent perturbation series which is 
 quantitatively correct even for $T\simeq T_{K}$ and only fails to recover 
 the low $T$ Fermi liquid behavior, was proposed in Ref.\cite{Nagaoka65}. The 
 approach is based on a non-linear integral equation for the $t$-matrix for 
 non spin flip scattering which determines the one particle Green's function: 
 \begin{equation} 
 \begin{split} 
 \mathcal{G}(\mathbf{p,p^{\prime }};\omega _{n})& =\mathcal{G}_{0}(\mathbf{p}%
 ;\omega _{n})\delta (\mathbf{p-p}^{\prime }) \\ 
 & +x_{r}J\mathcal{G}_{0}(\mathbf{p};\omega _{n})t(\omega _{n})\mathcal{G}%
 _{0}(\mathbf{p}^{\prime };\omega _{n}), 
 \end{split} 
 \label{SC1} 
 \end{equation}%
 where $\mathcal{G}_{0}(\mathbf{p};\omega _{n})=1/(i\omega _{n}-\varepsilon _{%
 \mathbf{k}}+\mu )$ is the bare valence hole Green's function. $%
 x_{r}=x-x^{\ast }$ is the concentration of the degenerate impurities. M\"{u}%
 ller-Hartmann and Zittartz\cite{Zittartz70} solved the non-linear integral 
 equation for $t\left( \omega \right) $ exactly. The approach was applied to 
 study spin Kondo impurities in a superconducting host. A rich behavior for $%
 T_{c}\left( x\right) $ was obtained which was shown to agree well with 
 experiments\cite{Maple}. In what follows we use and generalize this approach 
 to investigate superconductivity in the charge Kondo problem. This 
 scattering matrix approach is unique as it allows to investigate the subtle 
 crossover close to $T_{K}$ and, as we will see, naturally includes effects 
 related to the coupling between impurities, $I^{\pm }\left( R\right) $, 
 effects which are very hard to include in other, more modern approaches to 
 the Kondo problem\cite{Read}. 
 
 In the normal state $t\left( \omega \right) $ of the charge and spin Kondo 
 problems turn out to be identical and we can simply use the results of \ 
 Ref. \cite{Zittartz70}. In the superconducting state an anomalous scattering 
 matrix, $t_{\Delta }\left( \omega \right) $, occurs. Superconductivity and 
 charge Kondo dynamics are much closer intertwined than in the magnetic 
 problem and determining $t_{\Delta }\left( \omega \right) $ becomes a 
 considerably more complex task. However, for the linearized gap equation 
 which determines $T_{c}$, $t_{\Delta }\left( \omega \right) $ is small and 
 progress can be made analytically. We obtain for small superconducting gap, $\Delta$
 \begin{equation} 
 t_{\Delta }\left( \omega _{n}\right) =t_{\Delta ,\mathrm{loc}}\left( \omega 
 _{n}\right) +t_{\Delta ,\mathrm{prox}}\left( \omega _{n}\right)  \label{tan} 
 \end{equation}%
 with contribution $t_{\Delta ,\mathrm{loc}}\left( \omega _{n}\right) =-\frac{%
 \Delta }{3}\left( \frac{t\left( \omega _{n}\right) }{i\omega _{n}}-\frac{2}{%
 V_{0}}\frac{dt\left( \omega _{n}\right) }{di\omega _{n}}\right) $ determined 
 solely by the local Kondo dynamics and a nonlocal, "proximity" contribution $%
 t_{\Delta ,\mathrm{prox}}\left( \omega _{n}\right) =-\frac{\left\langle 
 T^{+}\right\rangle \left( 1-2\pi i\rho _{F}Jt\left( \omega _{n}\right) 
 \right) }{2X_{n}}$ which is proportional to $\left\langle T^{+}\right\rangle 
 $, reflecting the broken symmetry at the impurity in the superconducting 
 state. We allow for a finite attractive $\mathrm{BCS}$-interaction, $V_{0}<0$%
 , of the host. $t\left( i\omega _{n}\right) $ is the normal state $t$-matrix 
 of Ref.\cite{Zittartz70} and $X_{n}=\rho _{\mathrm{F}}J\left( \psi \left( 
 \frac{1}{2}+n\right) -\psi \left( \frac{1}{2}\right) -\log \left( \frac{T_{K}%
 }{T}\right) \right) $ with digamma function $\psi \left( x\right) $. 
 Performing the usual disorder average\cite{Abrikosov60} we finally obtain a 
 linearized gap equation 
 \begin{equation} 
 \Delta =-V_{0}T\sum\limits_{\omega _{n},\mathbf{p}}\frac{\widetilde{\Delta }(%
 \widetilde{\omega }_{n})}{\widetilde{\omega }_{n}^{2}+\varepsilon _{\mathbf{p%
 }}^{2}},  \label{ling} 
 \end{equation}%
 where $i\widetilde{\omega }_{n}=i\omega _{n}+x_{r}\rho _{F}Jt\left( i%
 \widetilde{\omega }_{n}\right) $ and $\widetilde{\Delta }\left( \widetilde{%
 \omega }_{n}\right) =\Delta \left( 1+x_{r}\rho _{F}J\frac{t\left( \widetilde{%
 \omega }_{n}\right) }{i\left\vert \omega _{n}\right\vert }\right) +x_{r}\rho 
 _{F}Jt_{\Delta }\left( \widetilde{\omega }_{n}\right) $. $\left\langle 
 T^{+}\right\rangle $ is determined by the ability to polarize a static 
 pairing state at the impurity site, just like in the proximity effect in 
 superconductors or the RKKY interaction in the magnetic case. Close to $%
 T_{c} $, we find $\left\langle T^{+}\right\rangle =-\frac{J}{2V_{0}}\chi 
 \left( T_{c}\right) \Delta $ with local susceptibility of the Kondo problem, 
 $\chi \left( T\right) \propto \left( T+T_{K}\right) ^{-1}$. 
 
 We first consider the limit $V_{0}=0$, i.e. the host material is not 
 superconducting on its own, like \textrm{PbTe}. Only the $t_{\Delta }$%
 -contributions which are proportional to $V_{0}^{-1}$ contribute to $%
 \widetilde{\Delta }\left( \widetilde{\omega }_{n}\right) $. At high 
 temperatures, $T_{c}\gg T_{\mathrm{K}}$, one easily finds that only $%
 t_{\Delta ,\mathrm{prox}}$ contributes to $T_{c}$ and we recover the mean 
 field result of Ref.\cite{Malshukov88}. The behavior changes as $T$ 
 approaches $T_{K}$. Now $\chi \left( T\right) \sim T_{K}^{-1}$ and $%
 t_{\Delta ,\mathrm{prox}}$ stops being the sole, dominant pairing source. 
 The pairing interaction becomes strongly frequency dependent. $t_{\mathrm{loc%
 }}\left( \omega \right) $ and $t_{\Delta ,\mathrm{prox}}\left( \omega 
 \right) $ become comparable to each other as well as to the pair breaking 
 scattering rate $\tau ^{-1}$ which is directly related to the existence of a 
 finite width, $\sim T_{K}$,\ \ of the Kondo resonance. \ Just like in case 
 of spin Kondo systems, pair breaking effects are largest for $T_{c}\simeq 
 T_{K}$. However unlike for the magnetic counter parts, the pairing 
 interaction itself \ strongly depends on $T_{c}/T_{K}$ and increases with 
 concentration.
 
\vskip -0.25cm
\begin{figure}[h]
\centerline{\psfig{file=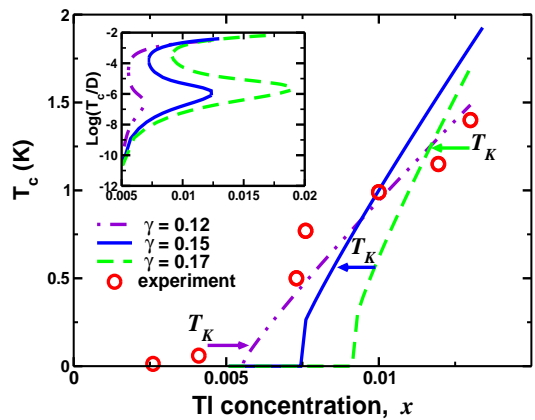,height=7.5cm,width=6cm,angle=-90}}
\caption{$T_{c}$ as a function of concentration for various values of the
dimensionless exchange coupling constant $\protect\gamma =\protect\rho _{F}J$%
. Experimental points\protect\cite{Fisher04} are plotted for comparison. 
Inset shows low-$T$ part of
concentration dependence of $T_{c}$ where re-entrance behavior appears.}
\label{Fig2}
\end{figure}

 Our results for the concentration dependence of $T_{c}$ are shown in Fig.2. 
 Charge Kondo impurities do indeed cause a superconducting state with 
 $T_c\simeq{T_K}$. At higher concentration we find $%
 T_{c}\ $rises almost linearly with $x$ whereas a rich behavior occurs in the 
 low temperature limit. The competition between pair breaking and pairing 
 interaction causes a reentrance normal state behavior which might serve as a 
 unique fingerprint for a charge Kondo origin of superconductivity. Due to 
 the uncertainty of the $\rho _{F}J$ value for \textrm{Tl}-doped \textrm{PbTe} 
 ~it is unclear whether this effect is observable in this material. In Fig. 2 
 we compare our results for several values of \ $\rho _{F}J$, chosen such 
 that $T_{K}$ $\simeq T_{c}$, with experiment\cite{Fisher04}. 
 To obtain $T_c\simeq{1}$K we used $D=\mu^{*}/4.5$ and 
 $\rho _{F}D\simeq{0.08}$. Given the above listed values for $%
 \rho _{F}\mu^{*}$ and $\mu^{*}$, these are perfectly reasonable 
 parameters, chiefly demonstrating that $T_{c}$ of several Kelvin is 
 possible within the charge Kondo theory for $x\approx{1}\%$. 
 These numbers further allow us to estimate the temperature $%
 \approx {30}$mK, below which the normal state reappears.  
\vskip -0.25cm
\begin{figure}[h]
\centerline{\psfig{file=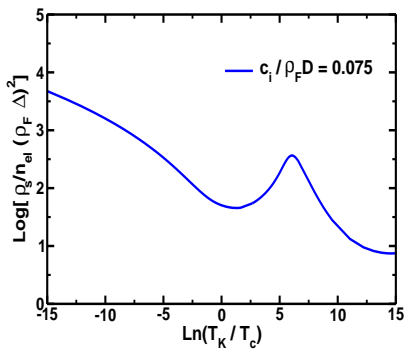,height=6cm,width=5.5cm,angle=-90}}
\caption{Normalized superfluid density as a function of $T_{K}/T_{c}$ shown
on a log scale. }
\label{Fig3}
\end{figure}
 Unlike ordinary superconductivity, pairing in charge Kondo systems is caused 
 by dilute impurities which are coupled by host carriers with low 
 concentration and the stability of the superconducting state with respect to 
 fluctuations becomes an important issue. In order to quantify this we 
 determine the superfluid density $\rho _{s}/n_{0}=\pi T\sum\limits_{\omega 
 _{n}}\widetilde{\Delta }^{2}(\widetilde{\omega }_{n})/\widetilde{\omega }%
 _{n}^{3}$ close to $T_{c}$, where $\rho _{s}\propto \Delta ^{2}$. In Fig. 3 
 we show our results for the dimensionless ratio $\alpha \equiv \frac{\rho 
 _{s}}{n_{0}}\left( \rho _{F}\Delta \right) ^{-2}$ as function of $%
 T_{K}/T_{c} $. \ $\alpha $ \ has a local minimum for $T_{c}\simeq T_{K}$, 
 caused by the strong scattering rate of a charge Kondo impurity which reduce 
 $\rho _{s}$. From $\alpha $ we can estimate the temperature, where phase 
 fluctuations affect the transition significantly and find that for $%
 T_{c}\simeq {T_{K}}$ superconductivity is robust, whereas for $T_{c}\ll {%
 T_{K}}$ the phase stiffness becomes rapidly small. In Ref.\cite{Malshukov88} 
 charge Kondo superconductivity was analyzed for $T_{c}\ll T_{K}$ with the 
 result that $T_{c}\simeq T_{K}\exp \left( -\lambda _{\mathrm{eff}%
 }^{-1}\right) $ and $\lambda _{\mathrm{eff}}\sim \frac{x}{\rho _{\mathrm{F}%
 }T_{\mathrm{K}}}$. Our result strongly suggest that this state is unstable 
 against phase fluctuations. 
 
 Within our theory we can also discuss the impact of charge Kondo impurities 
 in a system which is superconducting for $x=0$. We find, in agreement with 
 the quantum Monte Carlo simulations\cite{Schuettler89}, that $T_{c}$ 
 increases. Independent on $J$, $x$ is pair stabilization due to negative $U$ 
 centers always more efficient than pair breaking. 
 
 In summary we have developed a theory for superconductivity in charge Kondo 
 systems valid in the crossover region where $T\simeq T_{\mathrm{K}}$ which 
 can explain the comparatively large transition temperature in \textrm{Tl}%
 -doped \textrm{PbTe}. We showed that \textrm{Tl }is a very special impurity 
 as it first supplies a certain amount of charge carriers \ to the \textrm{%
 PbTe}-valence band and then puts itself into a self-tuned resonant state to 
 supply a new mechanism for superconductivity of these carriers. The subtle 
 interplay of pair formation and pair breaking by the same impurities can 
 cause a rich behavior including an enhancement of the host transition 
 temperature by impurities, a reentrance normal state transition and large 
 phase fluctuations of weakly coupled local pairs for $T_{c}\ll T_{K}$.
 Our results agree in order of magnitude and generic concentration dependence of $%
 T_{c}$ and $n_{0}$ with the experiments\cite{Fisher04,Murakami96,Nemov98} 
 for Pb$_{1-x}$Tl$_{x}$Te, strongly suggesting a charge Kondo origin for 
 superconductivity in this material. 
 
 We are very grateful for many stimulating discussions with Y. Matsushita, T. 
 H. Geballe, and I R. Fisher. This research was supported by Ames Laboratory, 
 operated for the U.S. Department of Energy by Iowa State University under 
 Contract No. W-7405-Eng-82 and a Fellowship of the Institute for Complex 
 Adaptive Matter (M.D.).

 \bigskip 
 

\begin{thebibliography}{99}
\bibitem{bandstructure} Following Refs.\cite{Nimtz83,Nemov98} we include the 
 bands around the $L$ and $\Sigma $-points of the Brillouine zone.
\end{thebibliography}
 \end{document}